\documentclass[12pt,preprint]{aastex}
\shorttitle{A new feature in the HB of NGC~6752}
\shortauthors{Momany et al.}
\begin{document}
\def\hst{{\sl HST}}
\def\farcm{\hbox{$.\mkern-4mu^\prime$}}
\def\farcs{\hbox{$.\!\!^{\prime\prime}$}}
\title{A NEW FEATURE ALONG THE EXTENDED BLUE HORIZONTAL BRANCH OF 
NGC~6752\footnote{Based on observations with the 
ESO/MPI 2.2m, located at the La Silla Observatory, Chile}}

\author{Yazan Momany\footnote{Dipartimento di Astronomia, Universit\`a
di  Padova,   Vicolo  dell'Osservatorio  2, I-35122     Padova, Italy;
momany-piotto-bedin@pd.astro.it},  Giampaolo   Piotto$^2$,   Alejandra
Recio-Blanco\footnote{INAF - Vicolo dell'Osservatorio 2, I-35122 Padova,
Italy;   recio@pd.astro.it},    Luigi     R.\    Bedin$^2$,     Santi\
Cassisi\footnote{INAF - Osservatorio Astronomico di Collurania, Via M.
Maggini,   64100 Teramo,  Italy; cassisi@te.astro.it},  and  Giuseppe\
Bono\footnote{INAF - Osservatorio Astronomico di Roma, Via Frascati 33, 00040
Monte Porzio Catone, Italy; bono@mporzio.astro.it}}

\medskip

\begin{abstract}

In this letter  we report  on the detection   of a new feature  in the
complex structure  of  the  horizontal branch   (HB) of the   Galactic
globular cluster  NGC~6752.  In the  $U$ {\it vs.}  $(U-V)$ plane, the
HB shows a discontinuity  (``jump'') at $U-V\simeq-1.0$ (corresponding
$T_{\rm   e}\sim23,000$K).   This  ``second  $U$-jump'' adds    to the
$u$-jump identified by Grundahl et al.\  (1999) in a dozen of clusters
at $T_{\rm e}\sim11,500$K.  We show that  this new discontinuity might
be due to the combination of post zero age  HB evolution and diffusion
effects.  We  identify 11 AGB-manqu\`e   stars. The comparison between
post-HB  star counts and   evolutionary   lifetimes, as predicted   by
canonical  stellar models,  shows  good  agreement, at  variance  with
similar estimates for NGC 6752 available in the literature.
\end{abstract}

\keywords{Globular Clusters: individual (NGC~6752) ---
stars: evolution --- stars: horizontal-branch --- stars: low-mass}

\section{INTRODUCTION}

Although   the global properties of  horizontal   branch (HB) stars in
Galactic  globular clusters (GCs) are  rather  well known, our current
understanding of these objects is still challenged by several puzzling
features which came out from a number of observations.  Among these we
mention: (a) a non-monotonic correlation between the HB morphology and
metal   abundance, i.e.\ besides   metallicity (first parameter) there
must  be a ``second''    parameter  (Sandage \&   Wildey 1967),  or  a
combination  of various parameters (Fusi   Pecci et al.\ 1993),  which
determine  the observed   HB  morphologies;   (b)  the    HB  is   not
homogeneously populated, and, in  particular,  all the HBs  with  blue
tails (BT) show the  presence  of  gaps, i.e.\  regions  significantly
underpopulated   by stars (Sosin et al.\   1997, Ferraro et al.\ 1998;
Piotto et al.\ 1999); (c) a fraction of the HBs with BTs are populated
by the so called extreme (or extended)  HB (EHB) objects, i.e.\ hot HB
stars  reaching temperatures of $30,000$K  or more (D'Cruz et al 1996,
Brown  et  al.\  2001, B01), both   in metal-poor  and  in  metal-rich
clusters (Rich et al.\ 1997).  These objects are the GC counterpart of
the field blue subdwarf population (Newell 1973);  (d) all the HB with
BTs present a   jump,  i.e.\ a  discontinuity  around   $T_{\rm e}\sim
11,500$K in the  Str\"omgren $u$, $u-y$  (Grundhal et al.\ 1999,  G99)
and  Johnson  $U$,  $U-V$ (Bedin et   al.\  2000, B00) color magnitude
diagrams (CMD).  Moreover, in  some of the  BTs there is  evidence of:
(i) a discontinuity in the relative abundance  of heavy elements (Behr
et al.\ 1999), (ii) a discontinuity in  the surface gravities (Moehler
et  al.\ 2000),  and,  finally (iii)  a discontinuity   in the stellar
rotation  velocities (Behr et al.\  2000,  Recio-Blanco et al.\ 2002).
The abundance anomalies, the  $u$-jump,  and the discontinuity  in the
(log $g$, log $T_{\rm e}$)   plane have been interpreted as  different
manifestations of  the same physical  phenomenon, i.e.\ the appearance
of  radiative metal  levitation (G99), though   Moehler et al.\ (2000)
point  out  that  this  mechanism  can only    partly account for  the
low-gravity problem.  Recio-Blanco et al.\ (2002) have also shown that
the discontinuity in the rotation velocity might be related to the G99
jump.

Considerable attention  (Landsman et al.\   1996, L96; Sweigart  1997;
D'Cruz et  al.\ 2000;  B01) has been   devoted also to  the EHB stars,
because  they  still represent  a  challenge  to canonical   HB models
(D'Cruz et al.\ 1996, B01), and also  because they are suspected to be
the  principal sources of  the  ultraviolet emission in the  so-called
$UV$-upturn galaxies (Greggio \& Renzini 1990).  To shed more light on
the origin of EHB stars, our group has  undertaken a long-term project
to   obtain  multiwavelength data  of  both  the   inner core  and the
outskirts of a number of EHB clusters.  Our main  goal is twofold: (i)
to constrain the main properties of EHB stars; and (ii) to investigate
the role of the  dynamical  evolution on the  origin of  these objects
(B00).

In this letter, we present preliminary results on one of the prototype
EHB clusters,    namely    NGC~6752.  This    object  is  a     nearby
[($m-M$)$_{\circ}=13.05$,   Renzini    et al.\   1996],  low reddening
[$E_{B-V}=0.04$,  Penny   \&  Dickens  1986], intermediate metallicity
([Fe/H]$=-1.64$) cluster, with   a  complex HB extending   to  $T_{\rm
e}\sim32,000$K.  We show   that the HB   of  NGC~6752 presents   a new
interesting feature, located at $T_{\rm e}\sim23,000$K, resembling the
G99 $u$-jump.   We    discuss whether current  theoretical   framework
accounts  for this ``second jump''.  We  also identify  post HB stars,
and compare their number with current evolutionary timescales.

\section{OBSERVATIONS AND DATA REDUCTIONS}

Images of  NGC~6752 through $U$, $B$,  and  $V$ filters  were taken on
July 25-26 2000, with the Wide-Field Imager (WFI) at the 2.2~m ESO-MPI
telescope  at  La  Silla, Chile.   The  WFI  camera consists  of eight
2048$\times$4096 EEV-CCDs, with a total  field of view of $34\times33$
arcmin$^2$. The exposure times  (30s and  150s in $U$,  5s and  10s in
$B$, and  3s and 7s in $V$)  were chosen in  order to sample  both the
bright   RGB and  the   faint MS  stars  .   Weather   conditions were
photometric,   with good seeing  (below   $0\farcs8$ for  all images).
Basic reductions   of the  CCD mosaic  was  performed  using the  IRAF
package  MSCRED (Valdes 1998),  while stellar photometry was performed
using  the DAOPHOT  and  ALLFRAME programs   (Stetson 1994).  Finally,
instrumental magnitudes were  calibrated to the $UBV$ standard  system
on a  chip to chip  basis using a set  of standard  stars from Landolt
(1992).

\section{THE COLOR-MAGNITUDE DIAGRAMS}

Figure 1 shows the  ($V$, $B-V$) and  ($U$, $U-V$) CMDs for stars with
$0\farcm3$$<R<$$13\farcm9$, where  $R$ is the projected  distance from
the  cluster center.   All the  sequences  of the CMD,  from below the
turn-off  up to the RGB  tip are well  defined and populated.  The CMD
shows an extended HB, which spans more than 4.5 magnitudes in $V$, and
extends down  to $V=17.9$, i.e.\   $\sim1.5$ magnitudes below the  TO.
The HB does not show any new feature in  the $V$, $B-V$ plane (Fig.~1,
{\it left panel}),  apart from the gap at  $B-V$=0 and $V=14.2$ (Caloi
1999), and a  sudden increase   in  the photometric dispersion   below
$V=17$ which cannot be explained in terms of photometric errors alone.
The   HB   itself  is  not   homogeneously  populated,  with  a  clear
underabundance of stars in the interval 17.0 $\le V \le$ 15.7.

It is surely more  instructive to look  at the $U$, $U-V$ CMD (Fig.~1,
{\it right panel}; see  also the zoomed version  in Fig.~2, {\it upper
panel}).  In this  plane, the  distribution of HB  stars is remarkably
discontinuous. In particular: (i) We note a  {\it jump} in the stellar
distribution at  $U-V\sim-0.30$.  This jump is  analogous to  the {\it
jump} identified by B00 in the same bands in NGC~2808, and corresponds
to the G99 jump.  (ii) Most interestingly,  at $U\sim 15.9$, $U-V \sim
-1.0$  there is a  ``second  jump'' in the   distribution of the stars
along   the HB.  This  feature  has never been   identified, though, a
posteriori, a similar jump can be seen in the CMDs of M13 and NGC~6752
in G99, and, to a lesser extent, in  the CMD of NGC~2808  of B00.  The
larger  photometric dispersion and  the smaller sample  of G99 and B00
partly  mask  this feature  in their  CMDs.  (iii)  There are 13 stars
significantly bluer and brighter than the bulk of HB stars (plotted as
open circles in  Fig.~2,  {\it lower panel}).  Most  of them  are  the
progeny of EHB stars.

Before turning our attention to the  newly identified ``second jump'',
we briefly comment on the post-HB candidates.

The evolution  of extreme  HB stars  and  their progeny have  received
special  attention  from  the community,   since they  are crucial  to
understanding  the Spectral Energy Distribution   (SED) of hot stellar
populations in    metal-poor and    metal-rich stellar   systems.   In
particular, we     are interested   in   testing   whether   predicted
evolutionary lifetimes are supported by empirical data since they have
a significant impact on the SED  of hot stellar populations.  Previous
investigation of NGC~6752 from UIT images (L96) showed a deficiency of
a factor of 3 in the observed number of post-EHB stars with respect to
the theoretical  predictions.   In  Fig.~2    there are  13    post-HB
candidates.  Four stars ({\it  circles} in Fig.~2, {\it  upper panel})
have been classified   as  post-EHB (AGB-manqu\`e) stars by   Moehler,
Landsman,  \& Napiwotzki (1998).  Note  that the  double open  circles
(stars B852 and B4380 in Moehler  et al.  1998) share almost identical
magnitudes and colors  in our catalog.  A  fifth post-HB candidate has
been proposed by L96  (UIT1 in L96,  {\it open square} in Fig.~2), and
we confirm their result.

Figure~2  ({\it lower  panel}) shows  the  comparison in the  $U, U-V$
plane between the observed   CMD and He-burning  evolutionary  models.
The thick line shows  the Zero-Age-Horizontal-Branch (ZAHB), while the
thin lines display the off ZAHB evolution of selected AGB-manqu\`e and
Post-Early-AGB structures (Greggio  \&  Renzini  1990).   Evolutionary
computations have been performed by  adopting an initial He content of
$Y$=0.23 and a metallicity $Z$=0.0006.   The ZAHB models were computed
by adopting the He core mass and He  surface abundance given by an RGB
progenitor  with mass  equal  to $0.8$M$_\odot$.    These  models were
constructed by  neglecting atomic  diffusion and radiative  levitation
(see Cassisi \& Salaris 1997 for more details).  The transformation to
the  observational plane  was     done  by adopting    the  bolometric
corrections and the color-temperature relations by Castelli, Gratton,
\& Kurucz (1997).  The comparison of the  observed CMD with the set of
models in Fig.~2 allows us to isolate the HB stars that, after the end
of  the   central He-burning  phase,     will eventually  evolve  into
AGB-manqu\`e structures.  The  M$=0.520$M$_\odot$ model is the coolest
model able to  produce AGB-manqu\`e stars. In Fig.~2,  there are 84 HB
stars hotter than this  track; only 11 of the  13 post-HB stars can be
classified as AGB-manqu\`e.  The ratio  between the two populations is
0.13, and    it is consistent,   within  the  uncertainties,  with the
predicted value  of  $\sim0.09$, thus solving the  discrepancy pointed
out by L96.  The 11  post-HB stars have been  inspected by eye, one by
one. Apart from one case, there are no obvious blends or CCD artifacts
that  could cause large photometric errors.   We believe that the much
higher resolution of our data can explain the difference in the number
of post-HB stars identified in the two works.

\section{THE SECOND JUMP}

The most intriguing  feature  in the  CMDs of Figs.~1   and 2  is  the
presence  of the ``second  jump''  around  $U-V\approx-1.0$.  The  two
arrows in Fig.~2 mark the position of the two  jumps along the HB blue
tail.  The comparison  with the models   in Fig.~2 shows  that the two
jumps  are   located     at $T_{\rm     e}\sim11,600$   and    $T_{\rm
e}\sim23,000$K. The cooler  part of the  CMD is well reproduced by the
models,     but HB   stars   hotter   than    $T_{\rm e}>11,600$K  are
systematically brighter than  predicted by ZAHB models.  However,  the
observed HB  stars are steadily  approaching the theoretical ZAHB when
moving from  $T_{\rm  e}\sim11,600$K to $T_{\rm e}\sim23,000$K,  where
the ``second jump'' suddenly appears.

Figure~2  shows that  the HB  extends  to $U=16.8$, corresponding to a
temperature $T_{\rm e}\sim32,000$K.   We  can  therefore exclude   the
possibility that the ``second  jump''  is related  to the presence  of
late helium flashers with envelope  mixing  (B01), as these stars  are
expected to be hotter than  $T_{\rm e}\sim32,000$K.  Another  possible
explanation for the origin of the ``second jump''  could be related to
the presence of  binaries  in the EHB  stars of  NGC~6752 (Peterson et
al.\ 2002).  Unresolved,  equal mass  binaries  would appear  brighter
than single stars with  the same temperature  and luminosity, and this
would mimic  the  observed  ``second  jump''.   However, the  ``second
jump'' in the HB would imply a rather {\it ad hoc} binary distribution
along the HB.

Figures~2 and 3 show  that the evolutionary  path of  hot HB stars  is
almost vertical in the $U$, $U-V$ plane, while  the ZAHB is not. It is
interesting   to  investigate whether  the  ``second  jump'' can be an
artifact of post-HB evolution in  the photometric plane of Figs.~2 and
3.  To   this purpose, we have   calculated the post-ZAHB evolutionary
times for stars  hotter    than  $T_{\rm e}=23,000$K. We  find    that
He-burning  stars evolving from the ZAHB  should spend $40\%$ of their
time  at magnitudes $U>16.3$,  and about $50\%$ of  their time at 15.8
$\le U \le$ 16.3. The ratio between these two lifetimes is 0.8. On the
other hand, we count 15 stars in the {\it lower box} of Fig.~3, and 30
stars in the {\it upper box}, to  which we should  add 3 more stars on
the  blue side of  the {\it upper box},  that are  very likely objects
which started their post-ZAHB evolution from the  {\it lower box}. The
observed ratio  is $0.46\pm0.20$.  Therefore, the stellar distribution
in the hottest portion of the HB in NGC~6752 can not be due {\it only}
to  the  peculiar morphology of  evolutionary tracks   in the $U, U-V$
plane.  Moreover, if we interpret  all the stars in  the two boxes  of
Fig.~3 simply as HB stars evolving from the ZAHB  as expected from the
canonical models, we are faced with a further anomaly: a sharp peak in
the  distribution of stellar masses. According  to the models, all the
stars with  $T_{\rm e}>23,000$K have a  mass of 0.505$M_\odot$, with a
dispersion  of only 0.003$M_\odot$.  It is difficult to understand the
origin of such a peculiar distribution along the HB.

Since    the stellar distribution   across the  hotter ``second jump''
resembles the G99's jump, one might wonder whether both features share
similar physical  origins.   The  occurrence of   the  G99's jump  was
explained   as the  aftermath  of radiative   levitation that causes a
substantial  increase in  the metal  content  of the outermost layers.
This  finding confirms  earlier results  by Glaspey  et al.\ (1989) in
NGC~6752, and later investigations by Behr et  al.\ (1999, 2000) for a
few other clusters.   Both  groups find that hot  HB  stars present He
depletion and a remarkable  over-abundances of heavy elements (Fe, Ti,
N, P, etc.), when compared with the cluster abundances. In particular,
in NGC~6752, Glaspey et al.\ (1989) found an over-abundance of iron by
a factor of 50  (and He depletion) in   the star CL1083 ($T_{\rm e}  =
16,000$K).   Instead, no abundance  anomalies  were found in the  star
CL1007 with   $T_{\rm  e} = 10,000$K.    The  results  by  Moehler  et
al. (1997),  who  find strong  He underabundances also  for stars with
$T_{\rm e}>23,000$K, suggests that diffusion is present in the hottest
stars, and might be related to the ``second jump'' anomaly.

Radiative   levitation  and   diffusion   are   possible   after   the
disappearance  of the envelope  convective layers located across the H
and HeI  ionization regions, at  $T_{\rm e} \sim10,000-11,000$K (Caloi
1999, Sweigart 2000).  For  $T_{\rm e}> 11,000$K, radiative levitation
takes place in a thin radiative layer  located between the surface and
the  second He ionization  region  (HeII).  When moving toward  hotter
effective temperatures the  radiative layer becomes thinner, since the
increase in the effective temperature causes a systematic shift of the
HeII  region  toward the  surface (Sweigart  2000).  Canonical stellar
models predict that the  HeII convective region approaches the stellar
surface in  the ZAHB   structures    at $T_{\rm e}\sim23,000$K.     In
addition, as expected on theoretical  grounds (Vink \& Cassisi  2002),
the  mass loss efficiency  increases with effective temperature.  Mass
loss works  as a competing process to   diffusion (Michaud \& Charland
1986), and indeed it decreases the  efficiency of radiative levitation
in  producing large chemical over-abundances of  heavy elements.  As a
result, radiative  levitation  is less  and   less  effective in   the
effective  temperature range   $T_{\rm e}\sim11,000-23,000$K, and  the
stars should steadily  approach the  canonical ZAHB luminosity,   when
moving  toward hotter  HB stars.   This  evidence is supported  by the
observed HBs (Fig.~2).  At the same  time, the over-abundance of heavy
elements should  decrease as  well.   Unfortunately, we  still lack of
observational data to support the latter hypothesis.
For stars  with $T_{\rm  e}>23,000$K, as a  consequence  of the larger
surface gravity and longer  central  He-burning lifetime, one  expects
that the  atomic  diffusion  becomes    more  and more  efficient   in
decreasing  the envelope He abundance, implying   the quenching of the
HeII  convection.   At the   same time,  the  large   increase in  the
effective   temperature  should enhance  the  capability  of radiative
levitation in producing a chemical separation in the stellar envelope.
According to this  working scenario, HB stars  with $22,000 \le T_{\rm
e}\le     24,000$K should present  a change    in the surface chemical
composition  in  comparison to  hotter and cooler   structures.  It is
rather tempting to associate  the  ``second jump'' identified in  this
paper with this  discontinuity in  the  chemical composition.  It   is
possibly  the   combination of this   discontinuity  and  the off ZAHB
evolutionary path and timescales  which produces the observed peculiar
morphology of the hottest part of the HB in NGC~6752.

Even though this scenario is quite attractive, the results of Michaud,
Vauclair   \& Vauclair (1983,  M83) could  somewhat  challenge it.  By
modeling the change in the   chemical stratification caused by  atomic
diffusion and radiative  levitation, they found  that the He abundance
in the envelope of hot HB stars sharply decreases in a very short time
scale (see their Figs.   1 and  2).  This  effect  is due to  the high
efficiency of  atomic  diffusion.   In  particular, their computations
suggest that  for structures  located  at $T_{\rm e}\sim20,000$K,  the
HeII  convective zone   should disappear  after  $\approx10^4$yrs.  If
these   predictions are correct, then  our  proposed scenario is ruled
out.  However, in   the  same investigation,  M83  have  qualitatively
estimated also the effects of turbulent motions, and found that if the
convective transport   is  particularly efficient it could   limit the
efficiency of atomic diffusion.   This means that  He, and in turn the
convective region  in the  envelope,  should not  rapidly disappear in
extreme HB structures.  This evidence is further supported by the fact
that HB  models constructed by  M83  which neglect  both the mass loss
(see Unglaub \& Bues 2001 for  a detailed discussion), and the effects
of turbulence, predict He abundances lower than observed.

In order  to estimate the minimum He  abundance  needed in the stellar
envelope of hot HB   stars to have  a  HeII convection zone, we   have
computed  several models  for  different assumptions   on the external
abundance  of  He.   We  have  found    that for  a   He abundance  of
$Y\approx0.1$, the envelope of extreme HB  structures still presents a
very thin convective layer across the HeII region.  On the other hand,
the spectroscopic   investigation by Moehler   et al.\ (2000) suggests
that the He  content in the EHB  stars of NGC~6752 with $17,000<T_{\rm
e}<22,000$K is   lower:   $0.01<Y<0.08$,   with  an  uncertainty    of
$\sim0.02$.  However, to reach firm conclusions concerning the cut-off
in the He   abundance for the  occurrence  of a convective region  one
needs  a  new  generation  of  evolutionary  models that  consistently
account   for  all   physical mechanism(s)    affecting  the  chemical
stratification of HB stars.

Nevertheless,  if the  suggested  scenario  is  actually at  work, the
presence  of a convective region  (very close to   to the surface in a
narrow mass  range in the  EHB) would imply that  the stars across the
``second  jump'', when compared with  the stars across the first jump,
should show larger macroturbulence  velocities and a larger spread  in
chemical abundances.  This qualitative  scenario is  somehow supported
by spectroscopic measurements of field sdB  and sdO stars. As a matter
of fact,  there is evidence  that He abundance steadily increases when
moving from  sdB to sdO.  In  these stars it  has also been observed a
strong depletion  of  silicon  and  a  mild underabundance  of  Carbon
(Lamontagne, Wesemael,  \&  Fontaine 1987), while  other  elements are
enriched even with respect to solar composition (Heber  et al.  2002).
Accurate abundance measurements of He and heavy elements along the hot
portion of the  HB  in GCs, such  as   NGC~6752, could be  crucial  to
constrain the  theoretical framework.  At the  same  time, new optical
and UV  spectroscopic data are needed to  ascertain the occurrence and
efficiency of mass-loss among EHB stars (Heber et al. 2002).

\acknowledgments 
We thank the anonymous referee for his/her comments which helped to
improve the paper presentation. We also thank S. Moehler and
W. Landsman for the useful discussions.  YM recognizes the support of
the Jordanian Ministry of Education, AR recognizes the support of the
{\it INAF}.  This program was supported by the {\it ASI} and by the
{\it MIUR}-Cofin2002 grants.

\clearpage

\begin{figure}
\plotone{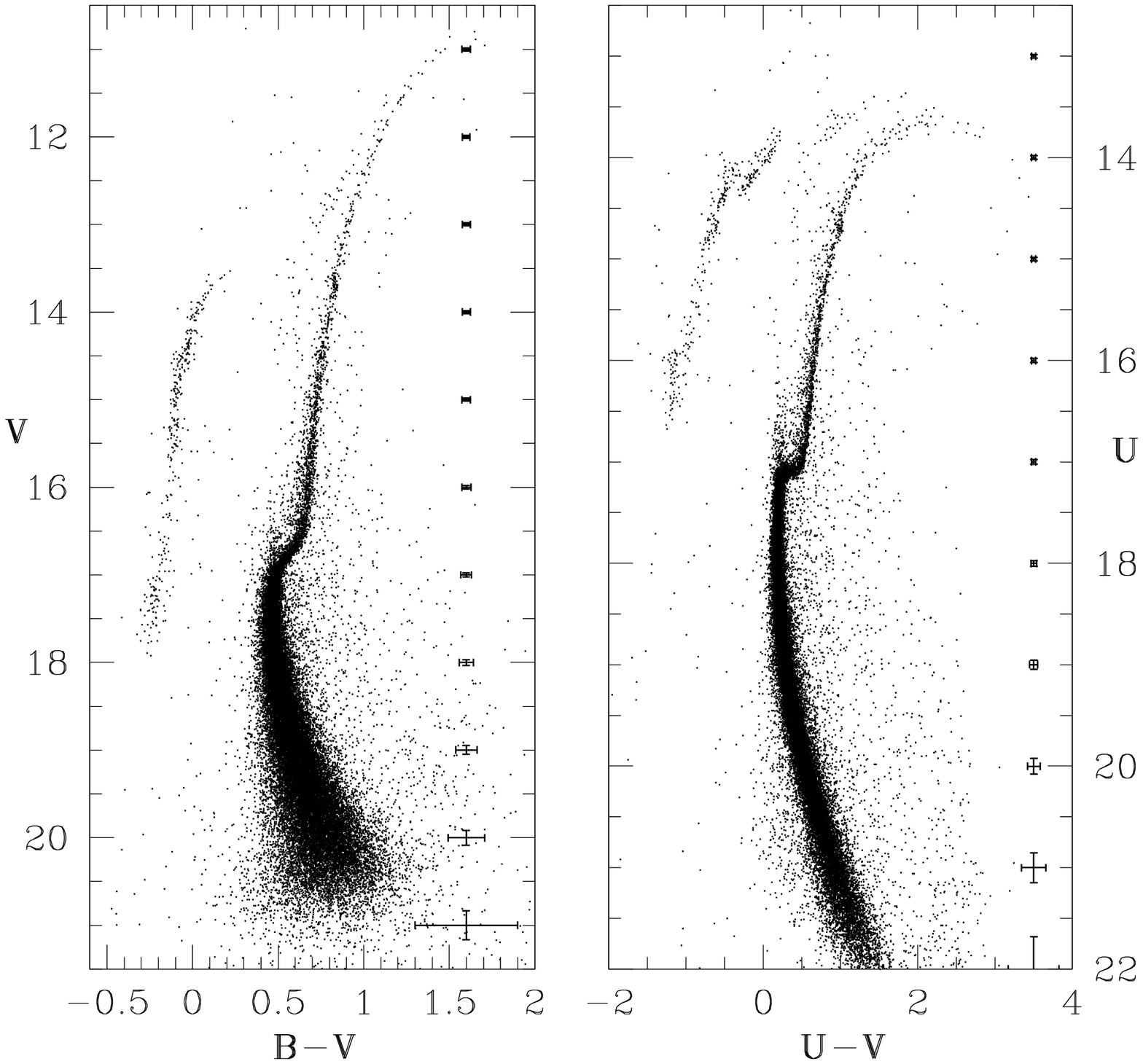}
\figcaption{The  ($V$,  $B-V$)  and  ($U$,
$U-V$) CMDs for  stars  with $0\farcm3<R<13\farcm9$, where $R$  is the
projected  distance from the cluster   center. Photometric errors,  as
calculated by ALLFRAME, are shown on the right side of the two CMDs.}
\end{figure}
\clearpage  
\begin{figure}
\plotone{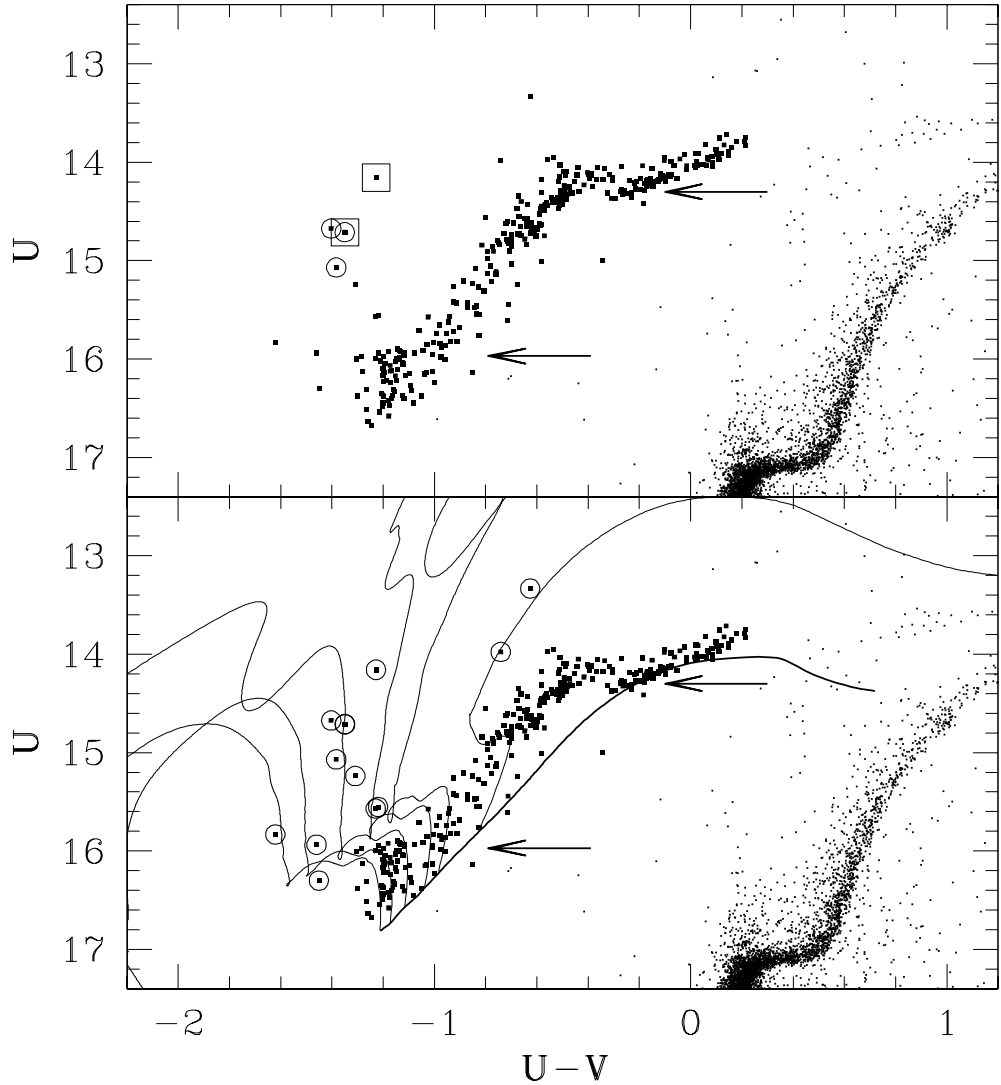}
\figcaption{Upper panel shows a close look
at the ``jumps''  in the HB  of  NGC~6752.  The four  open circles are
post-HB stars  from Moehler et al.\ (1998),  while the  open rectangle
shows the star  UIT1 of L96.  The  lower panel shows the  same diagram
with a canonical $Z=0.0006$  ZAHB model (thick  line).  Thin lines are
HB  models for stars with masses  M$=0.504$, $0.505$, $0.510$, $0.515$
$0.520$ and   M$=0.540$M$_\odot$.  Open circles  are   the 13  post-HB
candidates.}
\end{figure}
\clearpage  
\begin{figure}
\plotone{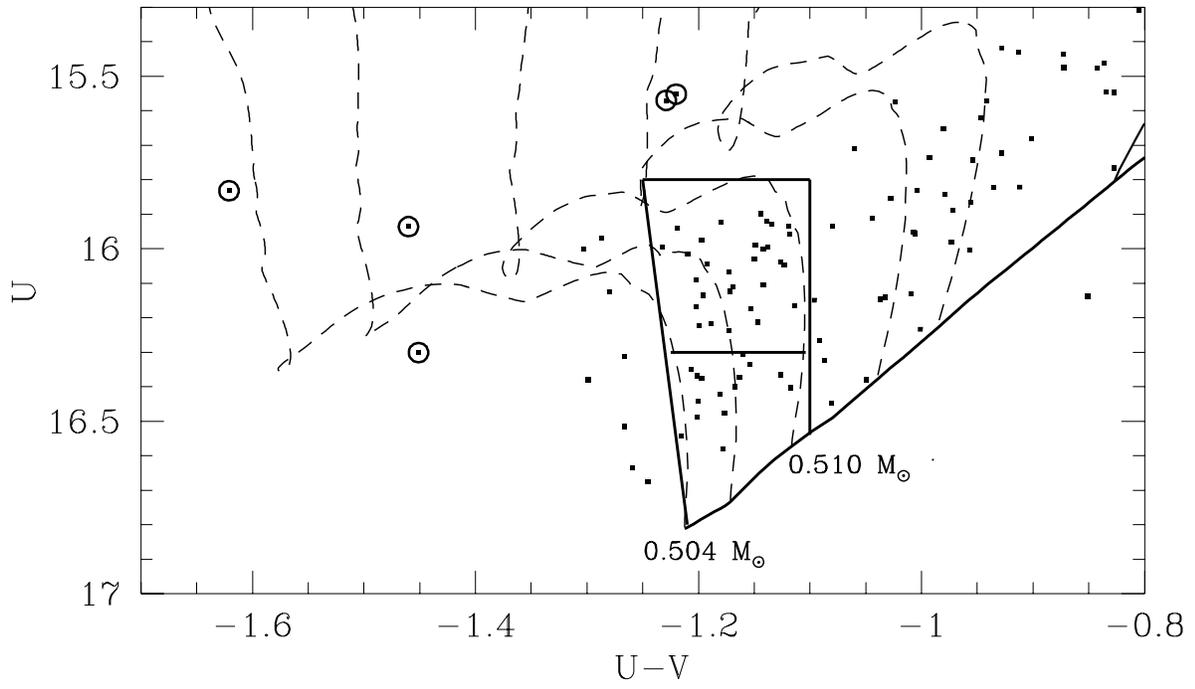}
\figcaption{As in Fig.~2, but zooming around the EHB.}
\end{figure}
\end{document}